# Density Formalism for Quantum Theory

Roderick I. Sutherland

Centre for Time, University of Sydney, NSW 2006 Australia

rod.sutherland@sydney.edu.au

**Abstract**

A simple mathematical extension of quantum theory is presented. As well as opening the possibility of alternative methods of calculation, the additional formalism implies a new physical interpretation of the standard theory by providing a picture of an external reality. The new formalism, developed first for the single-particle case, has the advantage of generalizing immediately to quantum field theory and to the description of relativistic phenomena such as particle creation and annihilation.

**1. Introduction**

The aim of this paper is to develop a possible generalization of the formalism of quantum theory. This generalization is suggested by some implications of Bell's theorem which follow from the assumed validity of special relativity and quantum mechanics. It takes the form of a mathematical description involving final as well as initial boundary conditions.

Since this model goes beyond just describing the results of measurements to portray an underlying reality, it also provides definite answers to some well-known interpretational questions raised by quantum mechanics. In contrast to the usual theory, the proposed model involves the continuous existence of various quantities in three-dimensional space (even in the case of an n-particle wavefunction which is only defined in configuration space of 3n dimensions). The quantities in question are densities, such as mass density, charge density, momentum density, energy density, etc., these all having definite values at every instant of time regardless of which observable is measured.

The structure of the paper is as follows. In Sec. 2, a discussion concerning Bell's theorem leads to the conclusion that maintaining the usual equivalence of all inertial reference frames necessarily implies the existence of retrocausality. This suggests the need to consider the notion of final boundary conditions. Sec. 3 gives this notion a specific mathematical form via the introduction of density quantities which are dependent on both initial and final conditions. It is then demonstrated in Sec. 4 that the new formalism is in agreement with the predictions of the standard theory. Since this extension of quantum theory involves a significantly different way of thinking, Secs. 5 is devoted to examining the new viewpoint in some detail. In particular, various thought experiments are discussed to highlight the main characteristics of the model.

We return to the mathematical development of the theory again in Sec. 6, where a derivation of the theory from a Lagrangian formalism is outlined. The next three sections are concerned with generalizing the model beyond the non-relativistic, single-particle case considered so far. In Sec. 7, the new formalism is extended to the many-particle case and then Sec. 8 generalizes it further to include creation and annihilation of particles. This last extension is achieved by postulating a simple rule expressible in terms of Feynman diagrams. The final generalization to incorporate relativistic quantum theory and quantum electrodynamics is formulated in Sec. 9.

Three short discussion sections are then presented. Sec. 10 briefly examines the nature of a decay process in this model and, in doing so, highlights the inherent smoothness of all such processes in the theory. Sec. 11 considers the question of testable consequences of the proposed formalism and outlines some new theoretical possibilities revealed by the additional framework provided. Sec. 12 then briefly discusses differences between this model and related theories. Finally the process of measurement is examined in detail in Sec. 13 and its continuous nature emphasised.

## 2. Motivation for final boundary conditions

The goal we are setting ourselves here is to formulate an extension of quantum mechanics which provides a picture of physical reality at all times, not just a description of measurement outcomes. The notion that an external reality does, in fact, exist in the absence of observation is the assumption of realism and will be adopted in what follows. As a first step towards such a theory, we will narrow the possibilities by assuming that we should work within the present four-dimensional space-time structure of special relativity. Some justification for this position lies in the fact that Lorentz invariance of the mathematical formalism is preserved in the transition from the classical to the quantum domain and there is no compelling evidence pointing to any particular modification or generalization.

Having said this, it is useful to begin by looking at some basic implications of Bell's theorem [1,2]. This well-known theorem concerns pairs of particles created with correlated spin states in certain decay processes. It refers to times when the particles have become widely separated in space and are apparently no longer interacting with each other, their spin correlation nevertheless remaining. The theorem implies that the statistical predictions of quantum mechanics require the existence of a nonlocal connection between such particles[1]. Specifically, Bell has shown that quantum mechanics is not compatible with the natural assumption that the result of a spin measurement on one particle of a correlated pair is independent of which spin component (if any) is measured on the other particle. This conclusion is somewhat perplexing, since the ad hoc introduction of signals passing across the space between the particles in order to accommodate Bell's result seems quite implausible.

The quantum mechanical predictions employed in the derivation of Bell's theorem have been well supported by subsequent experiments [2,3]. In addition, the experiments have established[3] that any proposed nonlocal communication must occur faster than light to be consistent with all the evidence[2]. More formally speaking, the emission and reception events for such a linkage would have to be in a spacelike relationship to each other. Now, since we are assuming the validity of special relativity, the time order of spacelike separated events is not absolute - it can be reversed simply by viewing them from an appropriate alternative frame of reference. This means that any such communication must occur backwards in time relative to some reference frames. Therefore, since all frames are on an equal footing, it then follows that special relativity and quantum mechanics taken together imply the existence of backwards-in-time effects, i.e., of "retrocausality".

---

[1] Some authors state that realism is also assumed in the derivation of Bell's theorem, but the relevance of this proposed further assumption is controversial and unclear.

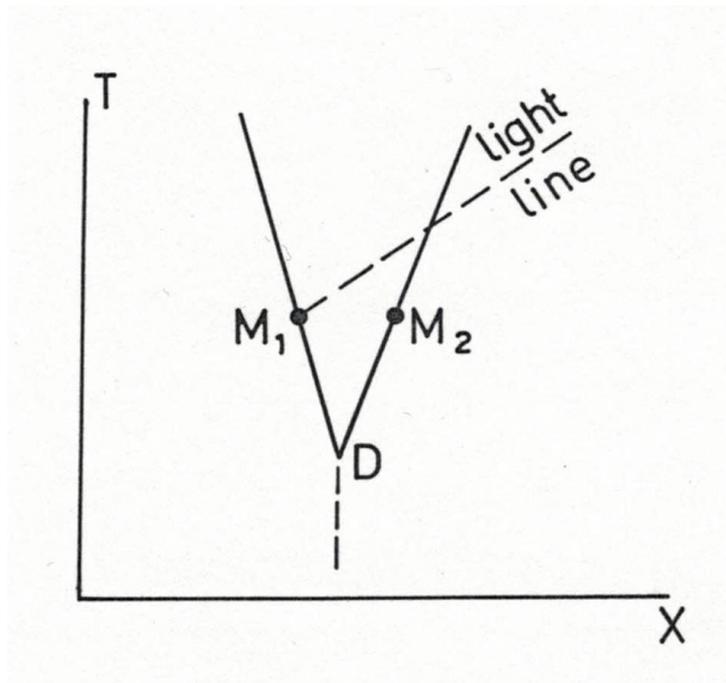

Fig. 1. Space-time diagram for the arrangement corresponding to Bell's Theorem. Two particles with correlated spin states are created at the decay point D and then measurements are performed on them at $M_1$ and $M_2$.

Pursuing this line of thought further, imagine that a spin measurement is performed on each particle of a correlated pair and that the two measurement events $M_1$ and $M_2$ are spacelike separated (see Fig. 1). Since we can change the time order of the two measurements simply by changing reference frames, the arrangement is symmetric in that neither measurement can be singled out as being the one which occurs first. Thus, if the outcome of the spin measurement on one particle is not independent of the direction chosen for the spin measurement on the other particle, we can conclude from the symmetry of the situation that any (hidden) communication causing this must be "two-way", i.e., each measurement must be capable of influencing the result of the other. Suppose the experimental arrangement is now adjusted by moving one of the two measurements (e.g., $M_2$) to a later time so that it is timelike ahead of the other. The key point to focus on here is that quantum mechanics makes the **same** statistical predictions for both spacelike and timelike separation of the measurements. This suggests that any communications, if they exist at all, will be the same in going to the timelike case, i.e., that they will remain "two-way". Now, one of these two directions must be backwards in time. Therefore it follows that, as well as being faced with retrocausal effects between spacelike separated events, we are also led to the notion of retrocausality between **timelike** separated events.

---

[2] It is, of course, not difficult to show that such a nonlocal effect could not be used for communicating information between observers.

One can, of course, avoid such retrocausal links by dropping the assumed equivalence of all reference frames and introducing a preferred frame in which the required communication between the particles propagates instantaneously. Here, however, we want to pursue the consequences of adhering to both special relativity and quantum mechanics, in accordance with all present experimental evidence.

A number of authors [4-11] have actually made use of the notion of retrocausal influences as a possible means of physically explaining Bell's mathematical result. Referring to Fig. 1 again, their models postulate the existence of a causal link along the path $M_1DM_2$. For example, the type of measurement performed on the first particle at $M_1$ is assumed to have a bearing on that particle's "state" at earlier times[3] (i.e., between $M_1$ and D). This in turn affects the other particle's state forwards in time from the decay point D, thereby affecting the result of the measurement at $M_2$. In a sense, the apparent action at a distance in three-dimensions then becomes a local connection when viewed from a four-dimensional viewpoint. One advantage which can be seen in such models is that they do not require communication along spacelike paths between the two particles through regions where nothing is expected to exist and where shielding could possibly be introduced to attempt to block any propagation.

The essential idea proposed in the above models is that the state of a particle at any time may be partly determined by the particle's future experiences as well as its past. Our aim in the present paper is to focus on such retrocausality and provide a consistent mathematical formalism which incorporates it. In doing so, we are led to a possible generalization of quantum theory, as well as a way of clarifying the theory's interpretation. It should be noted that, in considering this possibility, we are not suggesting the existence of any **movement** through four-dimensional space-time in either the forwards or backwards time directions. (This would require a fifth dimension to act as a "super time".) Motion remains confined to the three-dimensional picture.

As a first step towards developing such a formalism, we must deal with the question: what aspects of a particle's future are relevant? Possible factors could be the type of measurement to be performed next, the nature of the particle's interaction with the next particle it encounters and perhaps the nature of all future measurements and interactions. An indication of the best way to proceed is obtained by looking at the usual way we take account of a particle's **past** experiences: we work with an initial wavefunction $\psi_i$ which summarizes the particle's relevant past. More formally speaking, $\psi_i$ specifies the initial boundary conditions. It seems natural, therefore, to supplement $\psi_i$ with a "final" wavefunction $\psi_f$ specifying the **final** boundary conditions. As with $\psi_i$, the final wavefunction $\psi_f$ will be restricted to being a solution of the time-dependent Schrödinger equation[4].

---

[3] The word "state" as used here refers not only to the usual quantum mechanical description but also to any other quantities of which we are not presently aware.

The new wavefunction $\psi_f$ introduced here is independent of the usual wavefunction $\psi_i$ and should not be confused with the result of evolving $\psi_i$ deterministically to a later time. Thus, at any single time t, there are two distinct wavefunctions: (i) the initial wavefunction $\psi_i(\mathbf{x},t)$, which summarizes the initial boundary conditions existing at some earlier time $t_1$ and which has been evolved forwards from $t_1$ to t and (ii) the final wavefunction $\psi_f(\mathbf{x},t)$, which summarizes the final boundary conditions at some later time $t_2$ and which has been evolved back from $t_2$ to t. In the detailed model developed below on this basis, specifying $\psi_i$ at time $t_1$ and $\psi_f$ at time $t_2$ then determines what exists at any intermediate time.

**3. Density expressions**

Up to this point we have been talking loosely in terms of "particles". To proceed further, however, we need to address the question: exactly what sort of entities is the quantum mechanical formalism describing? As specified earlier, we are assuming that these entities reside in space-time. They must therefore be either localized, i.e., particles, or spread out, i.e., fields/waves. The mathematics of the Schrödinger formalism of quantum mechanics is of the form which one associates with fields and waves (with nothing indicating the existence of trajectories and world lines, for example). Hence one is tempted to propose fields as the underlying physical reality. On the other hand, a position measurement always yields a fully localized particle with a definite position value. This makes it seem natural that, e.g., electrons also have definite positions when we are **not** looking. In particular, the idea that an electron before measurement should be faithfully described by its wavefunction in being initially spread out and that, upon measurement, it should then collapse discontinuously to one point seems intuitively unlikely and unappealing[5]. Furthermore, such an instantaneous collapse is difficult to reconcile with relativity.

In deciding between these two possibilities, we will be influenced by the fact that the introduction of retrocausal effects adds a further perspective to the questions of interpretation. For example, as will be described below, it provides a reasonable way of escaping from discontinuous collapse of physically real entities. In addition, if we do not insist on a world line picture, a simple and elegant way of extending the standard quantum formalism to include the final wavefunction $\psi_f$ will be seen to suggest itself.

---

[4] Of course, the use of wavefunctions is only convenient in the non-relativistic approximation. The appropriate relativistic generalization of this scheme, involving propagators, is given in Secs. 8 and 9.

[5] In this regard it is also surprising that simple experimental set-ups for measuring position, such as the insertion of a photographic plate in the electron's path, can produce such a drastic change.

For these reasons we will choose here to pursue a field-like model where, by "field-like", it is meant that the usual quantities of interest, such as momentum, energy, mass and charge, will generally not be localized.

In accordance with this choice, we require a description of reality in terms of **densities** of the usual observables as functions of position **x**. (Here we are referring to smeared-out quantities, not probability distributions.) We already have expressions for such densities in the formalism of single-particle quantum mechanics, although with a somewhat different interpretation. In the non-relativistic (Schrödinger) case we have the following[6]:

$$
\begin{aligned}
&\text{Mass density:} \quad &\rho_m(\mathbf{x}) &= \frac{1}{N}\psi^*(\mathbf{x})\,m\,\psi(\mathbf{x}) \\
&\text{Charge density:} \quad &\rho_e(\mathbf{x}) &= \frac{1}{N}\psi^*(\mathbf{x})\,e\,\psi(\mathbf{x}) \\
&\text{Momentum density:} \quad &\mathbf{p}(\mathbf{x}) &= \frac{1}{N}\psi^*(\mathbf{x})\,\frac{\hbar}{2i}\overset{\leftrightarrow}{\nabla}\psi(\mathbf{x}) \\
&\text{Energy density:} \quad &E(\mathbf{x}) &= \frac{1}{N}\psi^*(\mathbf{x})\left[\frac{\hbar^2}{2m}\overset{\leftarrow}{\nabla}\cdot\overset{\rightarrow}{\nabla} + V(\mathbf{x})\right]\psi(\mathbf{x})
\end{aligned} \quad (1)
$$

Here, m and e are the total mass and charge of the particle concerned, the gradient operators $\overset{\rightarrow}{\nabla}$ and $\overset{\leftarrow}{\nabla}$ act to the right and left, respectively, $\overset{\leftrightarrow}{\nabla}$ stands for $\overset{\rightarrow}{\nabla} - \overset{\leftarrow}{\nabla}$, V(**x**) is an externally applied potential and N is a normalization constant given by

$$N = \int \psi^*(\mathbf{x})\psi(\mathbf{x})\,d^3x$$

this being equal to one when $\psi$ is normalized. (All integrations in this paper are from $-\infty$ to $+\infty$ unless otherwise specified.) Since the forms of the above expressions incorporate certain desirable characteristics (e.g., appropriate symmetries, conservation, simplicity, etc.) we will aim to construct our model via a minimal generalization of them.

The various densities are all bilinear in the wavefunction, i.e., they are of the general form

$$Q(\mathbf{x}) = \frac{1}{N}\psi^*(\mathbf{x})\,Q\,\psi(\mathbf{x}) \quad (2)$$

where Q(**x**) is the density of quantity Q and Q is the corresponding operator. It is relevant to our discussion to mention two other familiar densities here, both of which conform to Eq. (2). Associated with the charge density $\rho_e(\mathbf{x})$ above there is, of course, a current density of the form

---

[6] See, e.g., [12], Eq. (6.24), p. 134 and [13], Exercise 1.4, pp. 18 to 20.

$$\mathbf{j}(\mathbf{x}) = \frac{1}{N} \psi^*(\mathbf{x}) \, \mathbf{j} \, \psi(\mathbf{x})$$

Also, with regard to the momentum and energy densities, one can introduce a complete energy-momentum tensor

$$T^{uv}(\mathbf{x}) = \frac{1}{N} \psi^*(\mathbf{x}) \hat{T}^{uv} \psi(\mathbf{x})$$

of which $\mathbf{p}(\mathbf{x})$ and $E(\mathbf{x})$ are a part. (This will, in fact, be done for the relativistic case in Sec. 9.)

An important mathematical fact to note at this point is that the expressions for the current density and the energy-momentum tensor still satisfy the appropriate conservation laws[7] when they contain two **different** wavefunctions $\psi_1$ and $\psi_2$:

$$\mathbf{j}(\mathbf{x}) = \frac{1}{N} \psi_1^*(\mathbf{x}) \, \mathbf{j} \, \psi_2(\mathbf{x})$$

$$T^{uv}(\mathbf{x}) = \frac{1}{N} \psi_1^*(\mathbf{x}) \hat{T}^{uv} \psi_2(\mathbf{x})$$

Now, in our desired extension to quantum mechanics, we need density expressions which are functions of both the initial wavefunction $\psi_i$ and the final wavefunction $\psi_f$. Furthermore, these expressions should have zero values in spatial regions where either $\psi_i(\mathbf{x})$ or $\psi_f(\mathbf{x})$ is zero, suggesting that a product of the two is needed. The obvious way of proceeding is therefore to take expression (2) and replace $\psi$ with $\psi_i$ and $\psi^*$ with $\psi_f^*$, obtaining

$$Q(\mathbf{x}) = \frac{1}{A} \psi_f^*(\mathbf{x}) \, Q \, \psi_i(\mathbf{x}) \qquad . \tag{3}$$

The constant N is now written as A here because it has been transformed by these replacements into the **amplitude** connecting the initial and final states:

$$A = \int \psi_f^*(\mathbf{x}) \psi_i(\mathbf{x}) \, d^3x \qquad . \tag{4}$$

($\psi_i$ and $\psi_f$ will be taken to be separately normalized hereafter.) The choice of $\psi_f^* \psi_i$, rather than $\psi_i^* \psi_f$, has been made in both (3) and (4) to be consistent with the conventional notation of quantum theory[8].

Expression (3) is thus taken as the basic expression of our generalization of single-particle quantum mechanics and represents the way any physical quantity Q is smeared through space at any time. The various densities in this model all have definite values at

---

[7] i.e., they have zero 4-divergence in the case of relativistic quantum mechanics.
[8] See, e.g., [14], Eq. (5).

every instant of time, regardless of which quantity is measured. The particular densities of Eqs. (1) become

$$\text{Mass density:} \quad \rho_m(\mathbf{x}) = \frac{1}{A} \psi_f^*(\mathbf{x}) \, m \, \psi_i(\mathbf{x}) \tag{5}$$

$$\text{Charge density:} \quad \rho_e(\mathbf{x}) = \frac{1}{A} \psi_f^*(\mathbf{x}) \, e \, \psi_i(\mathbf{x}) \tag{6}$$

$$\text{Momentum density:} \quad \mathbf{p}(\mathbf{x}) = \frac{1}{A} \psi_f^*(\mathbf{x}) \frac{\hbar}{2i} \overleftrightarrow{\nabla} \psi_i(\mathbf{x}) \tag{7}$$

$$\text{Energy density:} \quad E(\mathbf{x}) = \frac{1}{A} \psi_f^*(\mathbf{x}) \left[ \frac{\hbar^2}{2m} \overleftarrow{\nabla} \cdot \overrightarrow{\nabla} + V(\mathbf{x}) \right] \psi_i(\mathbf{x}) \tag{8}$$

where A is given by (4) in each case. Also, any other observable quantity can be included in the model as a density via an expression analogous to those above. For example, spin is to be interpreted in this picture as an intrinsic angular momentum density spread through space (rather than as some sort of rotational motion of an underlying particle). It should be noted that this model differs somewhat from classical field theories such as Maxwell's electromagnetism in that the primary elements of reality here are not field amplitudes but densities, such as (5) to (8). (This will become more evident in the many-particle and relativistic cases.)

**4. Consistency with standard quantum mechanics**

To demonstrate consistency with the standard theory, we need to look at the new model's predictions relating to any quantity which is actually measured. A preliminary point to mention is that the change in a wavefunction upon measurement will be taken here as occurring discontinuously for convenience. This expedient assumption, however, is certainly not an essential feature of the model and a more thorough and physically realistic formulation of the measurement process is contained in Sec. 13, where all physical processes are found to evolve continuously.

Now, with regard to each of our densities, the quantity actually measured is the "total" value, corresponding to the integration of the density over all space. Suppose that a measurement of some observable quantity Q yields the eigenvalue q, with the subsequent state then being the eigenfunction corresponding to q. Consistency with standard quantum mechanics simply requires that a calculation in the new formalism of the total Q value after the measurement (via integration of the appropriate density expression) should give the measured value q. It is easy to show that this is the case. For example, consider a momentum measurement yielding the value **p**. The total momentum obtained by integrating the momentum density (7) over all space is

$$\frac{1}{A} \int \psi_f^*(\mathbf{x}) \frac{\hbar}{2i} \overleftrightarrow{\nabla} \psi_i(\mathbf{x}) \, d^3x$$

which, under integration by parts, becomes

$$\frac{1}{A} \int \psi_f^*(\mathbf{x}) \frac{\hbar}{i} \vec{\nabla} \psi_i(\mathbf{x}) \, d^3x$$

Now, in evaluating this expression for times after the measurement, $\psi_i$ must be taken to be the momentum eigenfunction which has arisen from the measurement. This eigenfunction satisfies

$$\frac{\hbar}{i} \vec{\nabla} \psi_i(\mathbf{x}) = \mathbf{p} \, \psi_i(\mathbf{x})$$

Hence, leaving the form of $\psi_f$ unknown and combining the last two equations, the total momentum can be written as

$$\frac{1}{A} \int \psi_f^*(\mathbf{x}) \mathbf{p} \, \psi_i(\mathbf{x}) \, d^3x$$

which, using Eq. (4) for A, reduces simply to $\mathbf{p}$, in agreement with the measurement outcome. The desired result has thus been obtained. Note that this derivation holds regardless of the form of $\psi_f$.

An analogous argument can be formulated to show that the total value **before** the measurement is also equal to the eigenvalue obtained. This will be the case assuming the final wavefunction $\psi_f$, propagating into the past away from the measurement, has become the eigenfunction corresponding to this eigenvalue. Looking at the momentum example again, integration by parts also allows the total momentum to be written in the form

$$\frac{1}{A} \int \psi_f^*(\mathbf{x}) i\hbar \overleftarrow{\nabla} \psi_i(\mathbf{x}) \, d^3x \tag{9}$$

and now $\psi_f$ satisfies the eigenvalue equation

$$\frac{\hbar}{i} \vec{\nabla} \psi_f(\mathbf{x}) = \mathbf{p} \, \psi_f(\mathbf{x})$$

which, under complex conjugation, can be written as

$$\psi_f^*(\mathbf{x}) i\hbar \overleftarrow{\nabla} = \mathbf{p} \, \psi_f^*(\mathbf{x}) \quad . \tag{10}$$

Combining Eqs. (4), (9) and (10) yields again the result $\mathbf{p}$ for the total momentum.

It is interesting that this last conclusion reintroduces the classical situation of the particular eigenvalue observed being in existence before the measurement is performed[9].

---

[9] This only applies to the observable actually chosen for measurement, not the alternative, unmeasured observables. Any other observable will have a pre-existing value too, but this value will not necessarily be the one "which would have been found". Also, position values are an exception because a position eigenfunction spreads out under both forwards and backwards time evolution and so the precise eigenvalue measured does not persist away from the measurement in either time direction.

Such a possibility is often thought to be excluded (except in the special case of prior preparation of the state) by the nature of the standard quantum mechanical formalism[10]. This feature fits naturally, however, once the theory includes retrocausality.

It is timely at this point to comment on the obvious fact that our density expressions are, in general, complex. This fact does not present any serious problem because the predictions of the theory corresponding to any **observed** values are real numbers, as the above consistency argument shows. In any case, it should be kept in mind that we also have the alternative of defining our general density to be the **real** part of our present expression instead, should any future considerations suggest this to be more appropriate.

**5. Further discussion of final wavefunctions and retrocausality**

To demonstrate that the various densities introduced in Sec. 3 really are retrocausally affected by future circumstances, consider two separate electrons each having the same initial wavefunction $\psi_i$ from time $t_1$ onwards. If we choose to perform measurements of different observables on the electrons at a later time $t_2$, they will have different final wavefunctions $\psi_f$ extending back from $t_2$ to $t_1$ (these being eigenfunctions of the respective observables measured). Since the general density expression is of the form $\psi_f^* Q \psi_i$ and so is obviously dependent on the final wavefunction, it follows that the values of all densities at any intermediate time between $t_1$ and $t_2$ will be different for the two electrons. Hence the type of measurement chosen at $t_2$ has a bearing on the physical reality existing at an earlier time. This example also indicates the way in which the initial notion of retrocausality has been given a specific mathematical form.

Note that it is not possible to interpret the $\psi_f$'s as instead arising at $t_1$, independent of the future measurements at $t_2$, and then propagating forwards in time. This is because it would then be inexplicable why, for each particle, the particular $\psi_f$ which arises at $t_1$ is certain to be an eigenfunction of the particular observable subsequently chosen at $t_2$.

We will now look at the nature of the final wavefunction $\psi_f$ in more detail. Since $\psi_f$, like $\psi_i$, evolves deterministically via the Schrödinger equation, specifying its form at any one time determines its form at other times. This means that if we knew both $\psi_i$ and $\psi_f$ at some initial time $t_1$, we could predict their values (and the values of all the densities) at later times. At first sight this seems to eliminate any retrocausality from the model again. The point is, though, that we cannot prepare and control the form of $\psi_f$ in the way that we can control $\psi_i$. Consider a simple experimental arrangement to illustrate this fact (Fig. 2). Electrons are emitted from a source and then are subsequently detected at another location. At some intermediate point between the source and the detector we insert some additional equipment to gain knowledge about the wavefunctions of the electrons.

---

[10] In particular, it is assumed to be excluded by impossibility proofs such as that of Kochen and Specker [15,16].

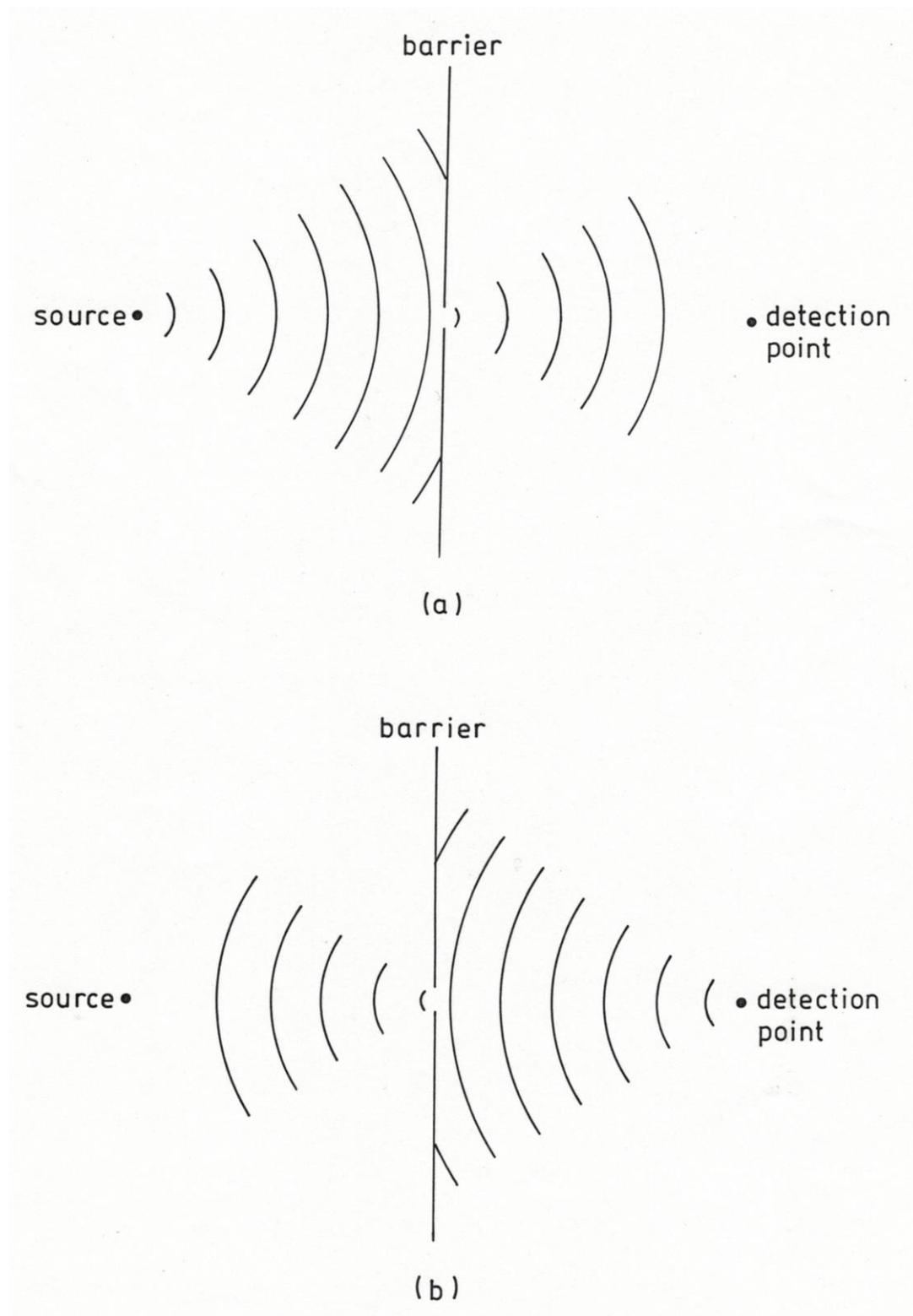

Fig. 2. Schematic representation of the spreading, and propagation through a slit, of (a) the initial wavefunction $\psi_i$ and (b) the final wavefunction $\psi_f$.

For simplicity we will choose the equipment to be a barrier containing a narrow slit and will consider only those electrons which pass through. The initial wavefunction will begin by spreading out gradually from the source, but will be reduced discontinuously to a narrow peak by passing through the slit. It will then gradually spread again as it continues on (see Fig. 2(a)). The final wavefunction, on the other hand, will gradually spread out as it propagates backwards in time from the final point of detection (Fig. 2(b)) then will be reduced discontinuously to a narrow peak as it strikes the back of the barrier and passes through the slit. It will then gradually spread again as it continues backwards in time towards the source.

Viewed in the forwards time direction, $\psi_f$ gradually reduces as it propagates away from the source and towards the front of the barrier, eventually becoming a narrow peak, all of which passes through the slit. Immediately after passing through, it discontinuously spreads out in a completely unpredictable way before propagating on towards the detector. This unpredictable behaviour demonstrates that we cannot control, or gain knowledge about, the future form of $\psi_f$ by such techniques (here, inserting a barrier containing a slit), since they can only provide information about the evolution of $\psi_f$ towards earlier times, its evolution towards later times being complicated by an apparently random change occurring immediately after our intervention. Hence the retrocausal nature of the theory cannot be argued away by claiming that we can determine the future form of $\psi_f$ via the initial physical conditions we impose. We can only determine what $\psi_f$ and the various densities have previously been, not what they are or will become.

It is worth reiterating here a point made earlier: in speaking of $\psi_f$ propagating into the past, we are not proposing the existence of any movement in four-dimensional space-time (motion being a three-dimensional phenomenon). Rather, we are merely saying that the future boundary conditions help determine the form of the present state (specifically, the various densities).

To illustrate other characteristics of the model, it is useful to consider two successive position measurements performed on an electron at times $t_1$ and $t_2$. The initial wavefunction $\psi_i$ starts as a delta function at $t_1$, then gradually spreads out as we move forwards in time towards $t_2$. The final wavefunction $\psi_f$, on the other hand, is a delta function at $t_2$ and gradually spreads out as we evolve it back in time towards $t_1$. Since all of the density expressions essentially contain a **product** of $\psi_i$ and $\psi_f^*$, these expressions have negligible values at points where either one of $\psi_i$ and $\psi_f^*$ is negligible. This means (see Fig. 3) that each density will initially (at $t_1$) be concentrated at one spatial point and will gradually expand, reaching a maximum width before contracting gradually and smoothly back to one point again at $t_2$. Unlike the usual discontinuous collapse of the wavefunction, the process is symmetric in time.

With the above picture in mind, it is worth referring back to the arguments considered in Sec. 3 in favour of particles always having definite positions. Those arguments were (i) that the discontinuous collapse of a smeared out electron down to one point when a position measurement is performed seems too unnatural a possibility and (ii) that such an instantaneous process is not easily reconciled with relativity. In answer we may say that, while our proposed model does entail spreading of the physically real quantities, it eliminates any discontinuous change in their spatial distribution at the time of a position measurement. Also, with regard to a relativistic treatment, Lorentz invariant versions of the various density expressions are easily constructed, as will be seen in Sec. 9.

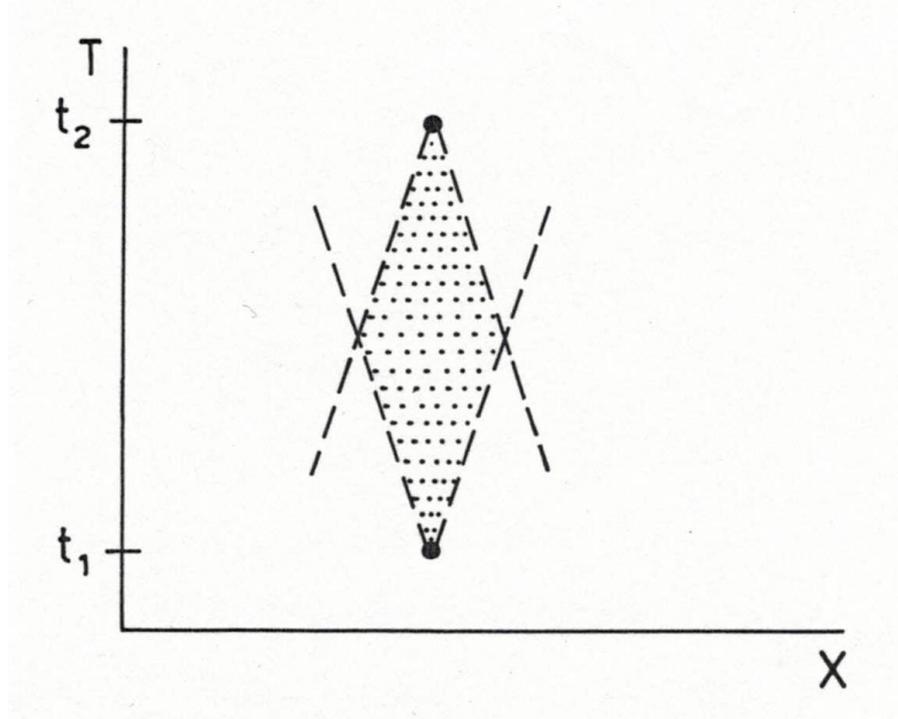

Fig. 3. Space-time diagram representing schematically the expansion and contraction of the density distribution between two successive position measurements (performed at times $t_1$ and $t_2$). The dotted region shows where the initial and final wavefunctions overlap significantly.

One can easily analyse other thought experiments along the same lines. For example, consider the familiar double-slit experiment: electrons travelling away from a source encounter a barrier containing either one or two open slits, with those that pass through eventually striking a detector. It is instructive to consider the various densities corresponding to those electrons which do pass through the barrier and to examine in what way they are varied by closing a slit. From the general form of the density expressions, one sees that the flow of mass, energy, charge, etc., away from the source will go towards (and through) both slits when both are open, but will only go towards one slit when one is open. This follows when one considers that closing a slit eliminates one of the two portions of $\psi_f$ propagating away from the barrier back towards the source and that all densities are negligible at points where is $\psi_f$ negligible.

**6. Lagrangian formalism**

It is possible to construct a Lagrangian formulation from which the proposed theory can be derived, as will now be discussed. The Lagrangian densities corresponding to the Schrödinger, Klein-Gordon, Maxwell and Dirac equations are all bilinear in the field. For example, the Schrödinger Lagrangian density is

$$\mathcal{L} = -\frac{\hbar^2}{2m}(\nabla\psi^*)\cdot(\nabla\psi) + \frac{i\hbar}{2}(\psi^*\frac{\partial\psi}{\partial t} - \frac{\partial\psi^*}{\partial t}\psi) - V\psi^*\psi$$

this being bilinear in $\psi$. By applying again our procedure of replacing $\psi$ with $\psi_i$ and $\psi^*$ with $\psi_f^*$, we obtain

$$\mathcal{L} = -\frac{\hbar^2}{2m}(\nabla\psi_f^*)\cdot(\nabla\psi_i) + \frac{i\hbar}{2}(\psi_f^*\frac{\partial\psi_i}{\partial t} - \frac{\partial\psi_f^*}{\partial t}\psi_i) - V\psi_f^*\psi_i$$

This new Lagrangian density can be taken as the basic axiom of our non-relativistic density formalism. From it we can deduce the field equations for $\psi_i$ and $\psi_f^*$ by varying first $\psi_f^*$ then $\psi_i$:

$$-\frac{\hbar^2}{2m}\nabla^2\psi_i + V\psi_i = i\hbar\frac{\partial\psi_i}{\partial t}$$

$$-\frac{\hbar^2}{2m}\nabla^2\psi_f^* + V\psi_f^* = -i\hbar\frac{\partial\psi_f^*}{\partial t}$$

Further, by employing Noether's theorem, we can determine the conserved quantities corresponding to this Lagrangian density and hence deduce expressions for the energy-momentum tensor and the charge and current densities. All our earlier density expressions are thereby obtained.

As a final point, it is perhaps worth mentioning that this formalism also offers an appealing way of viewing the bilinear form of Lagrangian densities for field theories: once the retrocausal aspect is properly incorporated, the various terms in each Lagrangian density contain products of the initial and final field functions and hence become **linear** in each field, this being the simplest type of dependence one can envisage.

**7. Many-particle case (non-relativistic)**

The density formalism can easily be extended to n-particle states. Recall that the single-particle version of our model in Sec. 3 was formulated by taking the usual density expressions of standard quantum theory and generalizing them appropriately to include the future boundary conditions. The same approach will be used for the many-particle case.

We will start with two particles and look at the usual expression for the density of some quantity Q. Consider the wavefunction $\psi(\mathbf{x}_1,\mathbf{x}_2)$ describing two distinguishable particles (e.g., an electron and a muon) in interaction, the particles having position coordinates $\mathbf{x}_1$ and $\mathbf{x}_2$, respectively. The density of Q for particle 1 has the form[11]

$$Q_1(\mathbf{x}) = \frac{1}{N} \int \psi^*(\mathbf{x}, \mathbf{x}_2) Q_\mathbf{x} \psi(\mathbf{x}, \mathbf{x}_2) \, d^3x_2$$

where the operator $Q_\mathbf{x}$ acts on the coordinates $\mathbf{x}$, not $\mathbf{x}_2$ (this operator being the same one as in the single particle case) and the normalization constant N is given by

$$N = \iint \psi^*(\mathbf{x}_1, \mathbf{x}_2) \psi(\mathbf{x}_1, \mathbf{x}_2) \, d^3x_1 \, d^3x_2$$

this being equal to one when $\psi$ is normalized. The analogous expression for particle 2 is

$$Q_2(\mathbf{x}) = \frac{1}{N} \int \psi^*(\mathbf{x}_1, \mathbf{x}) Q_\mathbf{x} \psi(\mathbf{x}_1, \mathbf{x}) \, d^3x_1$$

Note that in each case the coordinates of the other particle are integrated out. It follows that the total "Q density" for both particles together is

$$\begin{aligned} Q(\mathbf{x}) &= Q_1(\mathbf{x}) + Q_2(\mathbf{x}) \\ &= \frac{1}{N} \int \left[ \psi^*(\mathbf{x}, \mathbf{x}') Q_\mathbf{x} \psi(\mathbf{x}, \mathbf{x}') + \psi^*(\mathbf{x}', \mathbf{x}) Q_\mathbf{x} \psi(\mathbf{x}', \mathbf{x}) \right] d^3x' \end{aligned} \quad (11)$$

where $\mathbf{x}'$ is a dummy variable of integration.

Eq. (11) holds equally well for indistinguishable particles. In this case one only considers the total density $Q(\mathbf{x})$ (not the separate densities $Q_1(\mathbf{x})$ and $Q_2(\mathbf{x})$). Using the symmetry or antisymmetry of the wavefunction:

$$\psi(\mathbf{x}_1, \mathbf{x}_2) = \pm \psi(\mathbf{x}_2, \mathbf{x}_1)$$

the order of the coordinates in each $\psi$ in the second term of (11) can be reversed, simplifying the overall expression to

$$Q(\mathbf{x}) = \frac{2}{N} \int \psi^*(\mathbf{x}, \mathbf{x}') Q_\mathbf{x} \psi(\mathbf{x}, \mathbf{x}') \, d^3x'$$

For n indistinguishable particles this extends to

$$Q(\mathbf{x}) = \frac{n}{N} \iint \ldots \int \psi^*(\mathbf{x}, \mathbf{x}_1, \mathbf{x}_2, \ldots, \mathbf{x}_{n-1}) Q_\mathbf{x} \psi(\mathbf{x}, \mathbf{x}_1, \mathbf{x}_2, \ldots, \mathbf{x}_{n-1}) \, d^3x_1 \, d^3x_2 \ldots d^3x_{n-1}$$

(with an appropriate generalization of the normalization constant N). It is now a simple matter to generalize the above results to our density model by replacing $\psi$ and $\psi^*$ with

---

[11] See, e.g., [12], p. 338, Eq. (14.14) which, in conjunction with Eqs. (14.7) and (14.12), gives the current density of the i[th] particle in an n-particle system.

$\psi_i$ and $\psi_f{}^*$, respectively. Thus the density of some quantity Q in the two-particle case is given by the following:

(i) distinguishable particles

$$\text{particle 1:} \quad Q_1(\mathbf{x}) = \frac{1}{A} \int \psi_f{}^*(\mathbf{x},\mathbf{x}') Q_\mathbf{x} \psi_i(\mathbf{x},\mathbf{x}') \, d^3x' \tag{12}$$

$$\text{particle 2:} \quad Q_2(\mathbf{x}) = \frac{1}{A} \int \psi_f{}^*(\mathbf{x}',\mathbf{x}) Q_\mathbf{x} \psi_i(\mathbf{x}',\mathbf{x}) \, d^3x' \tag{13}$$

$$\text{total:} \quad Q(\mathbf{x}) = \frac{1}{A} \int \left[ \psi_f{}^*(\mathbf{x},\mathbf{x}') Q_\mathbf{x} \psi_i(\mathbf{x},\mathbf{x}') + \psi_f{}^*(\mathbf{x}',\mathbf{x}) Q_\mathbf{x} \psi_i(\mathbf{x}',\mathbf{x}) \right] d^3x' \tag{14}$$

(ii) indistinguishable particles

$$Q(\mathbf{x}) = \frac{2}{A} \int \psi_f{}^*(\mathbf{x},\mathbf{x}') Q_\mathbf{x} \psi_i(\mathbf{x},\mathbf{x}') \, d^3x' \quad . \tag{15}$$

In Eqs. (12) to (15), the normalization constant N has become the following amplitude:

$$A = \iint \psi_f{}^*(\mathbf{x}_1,\mathbf{x}_2) \, \psi_i(\mathbf{x}_1,\mathbf{x}_2) \, d^3x_1 \, d^3x_2$$

In extending to n distinguishable particles, an expression for the total density $Q(\mathbf{x})$ can be constructed in analogy to Eq. (14), the resulting expression having n terms instead of 2. For n **in**distinguishable particles, Eq. (15) becomes

$$Q(\mathbf{x}) = \frac{n}{A} \iint \ldots \int \psi_f{}^*(\mathbf{x},\mathbf{x}_1,\mathbf{x}_2,\ldots,\mathbf{x}_{n-1}) Q_\mathbf{x} \psi_i(\mathbf{x},\mathbf{x}_1,\mathbf{x}_2,\ldots,\mathbf{x}_{n-1}) \, d^3x_1 \, d^3x_2 \ldots d^3x_{n-1}$$

The form of A in the n particle case is

$$A = \iint \ldots \int \psi_f{}^*(\mathbf{x}_1,\mathbf{x}_2,\ldots,\mathbf{x}_n) \, \psi_i(\mathbf{x}_1,\mathbf{x}_2,\ldots,\mathbf{x}_n) \, d^3x_1 \, d^3x_2 \ldots d^3x_n$$

An obvious feature of this n-particle density formalism is that it provides a picture of a physical reality existing continuously in three-dimensional space. In contrast, the formalism of the standard theory is only defined in a mathematical space of 3n dimensions (wavefunctions being defined in configuration space) with the number of dimensions varying according to the number of particles involved.

The model also gives a clear picture of which entities can be considered separate and which are indivisible at any time. For instance, in the case of identical particles with single-particle wavefunctions overlapping in space, we can only speak of the **overall** density of any quantity at a point, whereas for non-identical particles there exists a separate density for each particle. This picture involving densities (rather than, say, underlying trajectories) also makes more understandable why indistinguishable "particles" have no individual identities.

## 8. Generalization to quantum field theory

In the high energy domain, where creation and annihilation of particles occurs, wavefunctions are no longer a convenient mode of description. Hence the present form of our density formalism cannot be directly generalized to incorporate such phenomena. An appropriate generalization follows quite naturally, however, once the theory is re-expressed in terms of propagators. This way forward will simply be postulated at first and then justified via a detailed example.

Consider a particular situation characterized by an initial state at one time and a final state at a later time. (For instance, this might be a particular scattering event specified by an initially prepared state and by the final results detected after the scattering has occurred.) The proposal here is that any density Q can be evaluated at intervening times by the following method. Referring to the set of Feynman diagrams describing the propagation of the system in position representation between the two times, the lines comprising each diagram should be "broken" one at a time and the appropriate single-particle operator $Q_\mathbf{x}$ inserted, as in Fig. 4. (This applies to external as well as internal lines.)

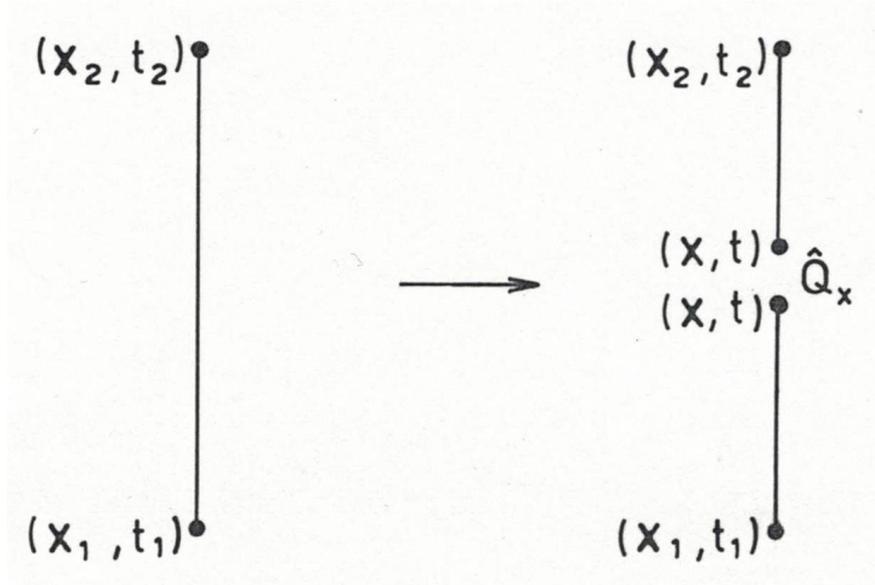

Fig. 4.  Procedure for obtaining densities from Feynman diagrams.

Algebraically this means that the propagator $K(\mathbf{x}_2, t_2; \mathbf{x}_1, t_1)$ corresponding to a line is replaced by[12]

$$\frac{1}{A} K(\mathbf{x}_2, t_2; \mathbf{x}, t) Q_\mathbf{x} K(\mathbf{x}, t; \mathbf{x}_1, t_1) \tag{16}$$

---

[12] The amplitude A connecting the initial and final states must also be included as shown in this equation, in analogy to Eq. (3).

Since each resulting diagram contains one broken line, the effect of this procedure is that each original diagram containing n lines is replaced by a series of n diagrams. The overall density is then found by summing the contributions of all the diagrams. **This procedure provides a direct way of generalising the model to quantum field theory.**

To provide justification for postulating the above rule, a derivation of the result (16) is provided in the Appendix for a simple case, namely the non-relativistic description of two non-interacting, identical fermions. In terms of diagrams, the amplitude for describing the propagation of such fermions consists of two terms which are conventionally represented by the Feynman diagrams in Fig. 5(a).

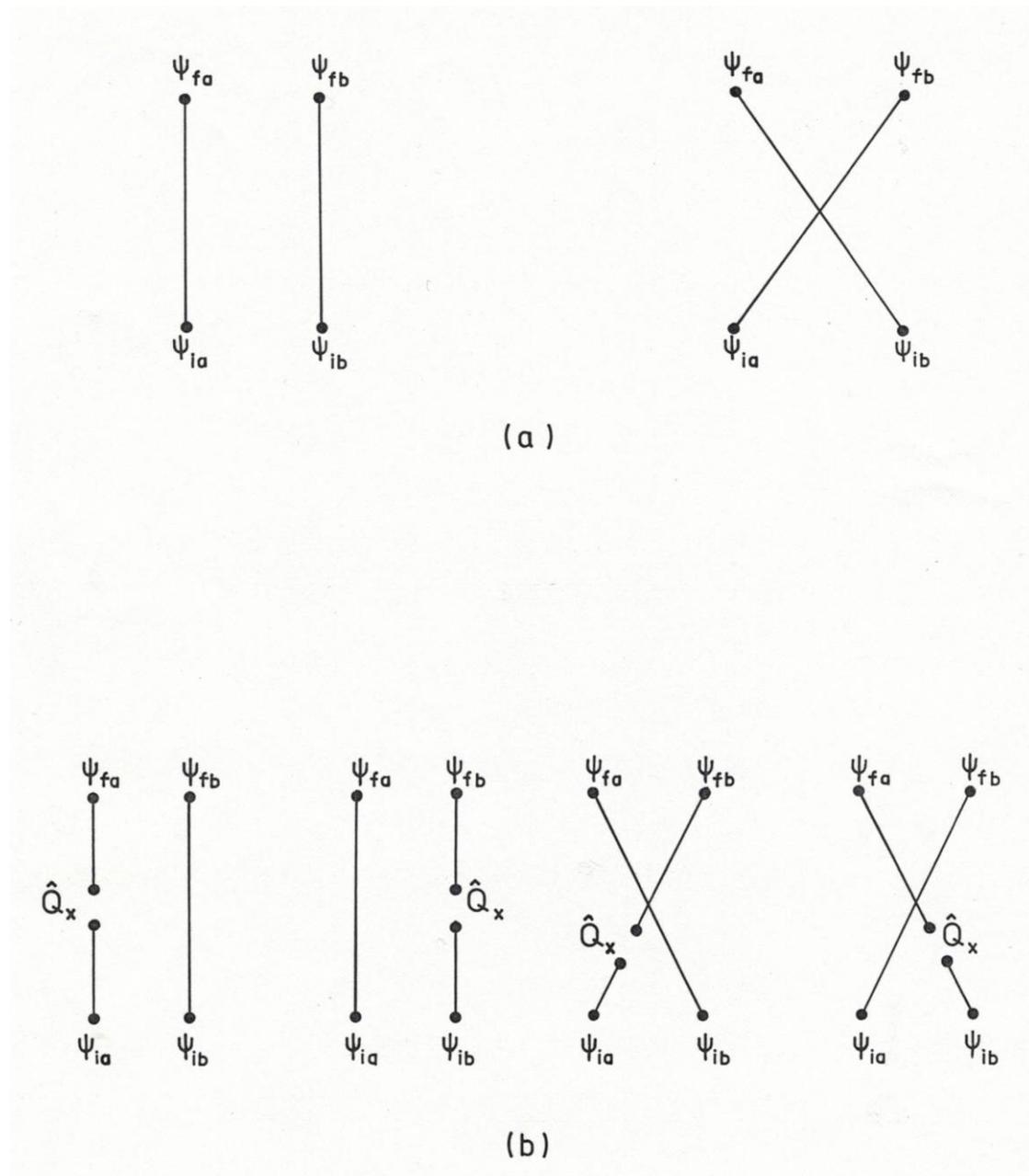

Fig. 5. (a) Feynman diagrams for a pair of non-interacting, identical fermions propagating from initial states $\psi_{ia}$ and $\psi_{ib}$ to final states $\psi_{fa}$ and $\psi_{fb}$. (b) Density diagrams corresponding to the process in (a).

The corresponding expression for the total density Q is then shown in the Appendix to be the sum of the four terms represented in Fig. 5(b). These four diagrams correspond to breaking one line at a time in the amplitude diagrams of Fig. 5(a) at some space-time point x. They thereby illustrate the procedure for evaluating any density, this method being directly generalizable to the relativistic domain where particle number is not conserved.

**9. Extension to relativistic quantum theory and quantum electrodynamics**

The relativistic generalization of our model will now be formulated. We will begin with the single-particle case of an electron described by a bispinor wavefunction $\psi$ satisfying the free space Dirac equation

$$\gamma^\mu \partial_\mu \psi + im\psi = 0 \qquad (17)$$

where we have set $\hbar = c = 1$. The gamma matrices satisfy

$$\gamma^\mu \gamma^\nu + \gamma^\nu \gamma^\mu = 2g^{\mu\nu} \qquad (\mu,\nu = 0,1,2,3)$$

with $\quad g^{\mu\nu} = \begin{pmatrix} +1 & & & \\ & -1 & & \\ & & -1 & \\ & & & -1 \end{pmatrix}$

The appropriate generalization of our non-relativistic density expression (3) to the single-particle Dirac case is simply

$$Q(x) = \frac{1}{A} \bar{\psi}_f(x) Q \psi_i(x) \qquad (18)$$

where $\bar{\psi} = \psi^\dagger \gamma^0$ is the adjoint to $\psi$, the dagger denoting Hermitean conjugation, and $x \equiv (x^0, \mathbf{x})$. The initial and final wavefunctions $\psi_i$ and $\psi_f$ are defined to be superpositions of positive energy eigenfunctions and the amplitude A is now given by

$$A = \int \bar{\psi}_f(x) \gamma^0 \psi_i(x) \, d^3x \qquad (19)$$

The particular densities which assume most significance in the relativistic domain are the 4-current density $j^\mu$ and the energy-momentum tensor $T^{\mu\nu}$ (other quantities of interest, such as charge density or energy density, being particular components of these). The operators for $j^\mu$ and $T^{\mu\nu}$ are[13]

$$\hat{j}^\mu = e\gamma^\mu$$

$$\hat{T}^{\mu\nu} = \frac{i}{4}\left[\gamma^\mu \overleftrightarrow{\partial}^\nu + \gamma^\nu \overleftrightarrow{\partial}^\mu\right]$$

---

[13] See, e.g., [17], p. 104, or [18], p. 419.

Inserting these operators into Eq. (18) to obtain expressions for the corresponding densities in terms of $\psi_i(x)$ and $\bar{\psi}_f(x)$, it can easily be shown (using (17) and its adjoint) that both these densities have zero 4-divergence, as required by the conservation laws for charge, energy and momentum.

To proceed towards a density formalism for quantum electrodynamics, we will use the Feynman propagator $K_+(x, x')$ to rewrite our equations in a more convenient form. This propagator satisfies the equation[14]

$$\gamma^\mu \partial_\mu K_+(x, x') + im K_+(x, x') = \delta^4(x - x')$$

and, for the time ordering $t_1 < t < t_2$, allows us to write $\psi_i$ and $\bar{\psi}_f$ in the forms

$$\psi_i(x) = \int K_+(x, x_1) \gamma^0 \psi_i(x_1) \, d^3x_1 \tag{20}$$

$$\bar{\psi}_f(x) = \int \bar{\psi}_f(x_2) \gamma^0 K_+(x_2, x) \, d^3x_2 \tag{21}$$

Now, in terms of $K_+$, the amplitude (19) connecting the states $\psi_i$ and $\psi_f$ can be written as

$$A = \iint \bar{\psi}_f(x_2) \gamma^0 K_+(x_2, x_1) \gamma^0 \psi_i(x_1) \, d^3x_1 \, d^3x_2 \tag{22}$$

Also, Eqs. (20) and (21) allow us to write the density expression (18) in the form

$$Q(x) = \frac{1}{A} \iint \bar{\psi}_f(x_2) \gamma^0 K_+(x_2, x) Q_x K_+(x, x_1) \gamma^0 \psi_i(x_1) \, d^3x_1 \, d^3x_2 \tag{23}$$

Comparing these last two equations it is clear that the density expression (23) corresponds to replacing the propagator $K_+(x_2, x_1)$ in the amplitude (22) by

$$\frac{1}{A} K_+(x_2, x) Q_x K_+(x, x_1)$$

This is, of course, in accordance with the rule formulated in Sec. 8 for extracting densities from Feynman diagrams: break the lines one at a time, insert the appropriate single-particle operator and divide by the overall amplitude (the required density then being the sum over the results from all diagrams). This rule provides the basis for extending our density formalism to quantum electrodynamics.

Having introduced the above rule it is actually more appropriate to adopt it, rather than our original expression (3), as the basic postulate of our theory. This is because, although we are taking the rule to apply in general, it is derivable from our earlier formalism only in the restricted case where there is no creation or annihilation of particles[15]. On the other hand, the earlier (wavefunction) formalism can all be derived directly from this propagator rule.

---

[14] See, e.g., [14], p. 752, or [17], p. 58.

**10. The nature of a decay process**

To appreciate further how the theory describes particle creation, it is instructive to look briefly at an example of a decay process. For convenience, the discussion in this section will be expressed in terms of initial and final states evolving through time. Consider the decay of an isolated neutron into a proton, electron and antineutrino. To describe this occurrence, we evolve the initial neutron state forwards in time and the final state describing the proton, electron and antineutrino backwards in time. The procedure formulated in earlier sections is then used to evaluate the densities in between. The initial neutron state will gradually develop a "proton-electron-antineutrino" amplitude in propagating forwards, whereas the final state will gradually develop a neutron amplitude in propagating backwards. The densities obtained by combining forwards and backwards neutron amplitudes will then gradually decrease in the forwards time direction, while the densities associated with the other three particles will gradually increase. All such processes in this theory thus consist of smooth flows between observations, with no discontinuous particle-like events occurring. This is despite the apparently sharp particle lines and vertices drawn in Feynman diagrams. These diagrams are a convenient aid in calculation, but one should be mindful that, in general, they simply represent terms in a perturbation series and one should not take the line and vertex picture too literally. In the neutron example, there is no single, precise time of decay in the density picture, the transition from neutron qualities to proton, etc., qualities occurring gradually.

**11. Possible tests of the theory**

The differences between this model and standard quantum theory relate principally to what exists at times **between** measurements. Therefore, at first sight the model appears to be not experimentally testable, since it seems to offer nothing new regarding the measurement results themselves. However, although the various densities are not directly measurable, their existence may be verified or disproved indirectly by testing other predictions which arise as a result of the additional theoretical framework provided.

The aim in this section is to indicate briefly some possible consequences of the density model that would allow it to be distinguished from the standard theory, thereby illustrating that the two theories are not merely equivalent. Two long-standing areas of concern immediately suggest themselves for re-examination from the perspective of the proposed new formalism, namely the procedure for evaluating the masses of elementary particles and the problem of constructing a quantum theory of gravity. In regard to particle masses, the theory offers a possible alternative approach for calculating them. Since the density formalism provides an expression for a particle's energy-momentum tensor $T^{\mu\nu}$ at any time, a value for the corresponding mass can be obtained by the classical method of integrating the energy density $T^{00}$ over all space. (An approximate value for $T^{00}$ can be obtained by using the Feynman diagram procedure described in Secs. 8 and 9 and summing over the lower order diagrams.)

---

[15] Of course, although our earlier formalism does not specifically require this rule to hold, it is the obvious generalization to make.

Turning to the issue of gravitation, the model allows the straightforward construction of a simple quantum gravity theory. In the Einstein field equations of general relativity:

$$G^{\mu\nu} = 8\pi T^{\mu\nu} \qquad (24)$$

the source term $T^{\mu\nu}$ represents the mass-energy distribution giving rise to the gravitational field. Normally, of course, this distribution is described classically. We may, however, generalize the theory to incorporate quantized mass-energy simply by turning again to the energy-momentum tensor of the density formalism and inserting it in Eq. (24). The resulting theory then treats matter via quantum mechanics, whilst the metric tensor contained in $G^{\mu\nu}$ continues to be treated as a classical field. Note, however, that this is more than just a semiclassical theory. The $T^{\mu\nu}$ of the present model is a precise, non-statistical quantity. It already incorporates the result of the next measurement because it contains the final wavefunction $\psi_f$ and so the probabilities of other outcomes are irrelevant. Since $T^{\mu\nu}$ is now definite, this means that $G^{\mu\nu}$ on the other side of the Einstein equation can also be "non-statistical" without introducing any inconsistency. The resulting theory can then yield definite (rather than "fuzzy") values for the curvature while remaining in full agreement with quantum mechanics.

Whether or not either of the above approaches turns out to be useful, they serve the purpose of showing that the density formalism is not merely equivalent to standard quantum theory.

## 12. Comparison with related theories

This section looks briefly at how the present model differs from some related theories. A consistent advocate of the retrocausality approach has been Costa de Beauregard [4]. His model can be regarded as a minimal extension of the orthodox theory in the sense that it maintains the Copenhagen interpretation. Costa de Beauregard departs from the usual description by including the extra feature that, once an eigenstate arises from measurement, that eigenstate then propagates both forward and backwards in time so that it exists prior to the measurement as well as subsequently. In contrast to our formulation, Costa de Beauregard does not feel any need for a model incorporating realism.

The transactional interpretation of Cramer [10] also involves backwards in time effects in that it contains advanced as well as retarded waves. The main point of difference is that the physical reality in Cramer's model corresponds to a **sum** of the retarded and advanced solutions, rather than a product. His model has the disadvantage that it does not provide a picture of physical reality located within three-dimensional space once we go beyond the single-particle case to a system of n entangled particles. Also, the model does not generalise in any obvious way to the case of quantum field theory.

The models of both Costa de Beauregard and Cramer do not contain the density expressions for mass, charge, energy, etc., which are characteristic of our model. Also, most of the remaining papers proposing retrocausality do not provide a mathematical formalism for describing this phenomenon in detail.

It is also worth mentioning for comparison another well-known model which maintains realism (but with no retrocausality), namely the hidden variables model of Bohm and de Broglie [19,20]. This model proposes that particles always follow definite trajectories. Compatibility with the predictions of quantum mechanics then entails the somewhat disquieting feature that particles must be subject to very non-local effects (perhaps under the influence of some accompanying field related to the wavefunction). Accepting this aspect for the moment, there are difficulties in generalizing the model of Bohm and de Broglie to the relativistic domain. In contrast, one of the main motivations behind the formulation of the density model is its ease of generalization to relativistic quantum field theory and such phenomena as particle creation and annihilation.

In the context of discussing the hypothesis of precise trajectories, it should be noted that the density model also provides an immediate explanation for the well-known fact [21,22] that the standard formalism of quantum mechanics does not single out any natural, non-negative expression for the joint probability distribution $P(\mathbf{x},\mathbf{p})$ for a particle's position $\mathbf{x}$ and momentum $\mathbf{p}$. Since the trajectory hypothesis requires each particle in an ensemble to have an underlying position value and momentum value at any time, one would expect that the formalism would be readily indicating such a joint distribution for these continuously existing values. In the density model, on the other hand, the smeared-out nature of a particle's position means that such a distribution is not applicable.

## 13. The measurement process

The aim of this paper has mainly been to suggest a possible physical reality which could exist between measurements in quantum theory. It has not been the primary intention to analyse the nature of the measurement process itself. In this regard, it was sufficient and convenient for the purposes of Secs. 4 and 5 to use the orthodox description of measurements in terms of discontinuous changes of the wavefunction, with a measurement treated as occurring at a single instant of time[16]. A more detailed and realistic description of the measurement process should, however, involve only **continuous** flows of the density quantities. Such a description will now be presented.

---

[16] Such a simplified description raises difficulties, as can be seen by considering the change in each density quantity **through** the time of measurement. For example, consider three different measurements performed in succession on an electron at times $t_1$, $t_2$ and $t_3$. The outcomes of the measurements will be taken to be the states $\psi_1$, $\psi_2$ and $\psi_3$, respectively. This means that the initial wavefunction of the electron will be $\psi_1$ between $t_1$ and $t_2$ and will change to $\psi_2$ between $t_2$ and $t_3$. Furthermore, the electron's final wavefunction will be $\psi_2$ between $t_1$ and $t_2$ and will change to $\psi_3$ between $t_2$ and $t_3$. Referring back to Eq. (3), the density of any quantity Q will therefore be of the form $\psi_2^* Q \psi_1$ just before time $t_2$ and will then change immediately to the form $\psi_3^* Q \psi_2$ afterwards. Such an instantaneous change in a spatially spread-out density would reintroduce the clash with relativity which our otherwise continuous picture has aimed to avoid.

An essential feature of any measurement is that it must allow us to distinguish between the possible outcomes and identify the result. This means that the possible states of the observed system (or of something with which it interacts) must become **separated in space**. The fact that all measurements share this general characteristic has been emphasized by other authors[17]. If the system is an electron, for example, this spatial separation may be the result of passing the electron's wavefunction through a magnetic field, or may correspond to scatter through different angles in a collision. Alternatively, after the electron's wavefunction interacts with the wavefunction of the relevant apparatus, the spatial separation could take the form of different pointer readings on a dial. This separation aspect of the measurement process takes place by continuous evolution via, e.g., the Schrödinger equation. To appreciate how it relates to our present problem, we will discuss a simple example and then use it as a basis for formulating the general case.

Consider a Stern-Gerlach set-up for measuring a particular spin component of a spin one-half particle. As the particle's initial wavefunction $\psi_i$ passes through the magnetic field of the apparatus, it is split gradually and continuously into two separate beams corresponding to the spin results $+$ and $-$. We will assume that the two beams remain separate and are not recombined (i.e., that the measurement is not "undone"). Despite the splitting of $\psi_i$ into two branches, it is an obvious experimental fact that measurements have single, definite outcomes. This definiteness is usually reconciled with the non-committal evolution of $\psi_i$ via the notion of wavefunction collapse.

In discussing this experiment, we need to keep in mind that the model formulated in this paper describes the underlying physical reality as consisting of densities of energy, mass, charge, etc. Hence the definiteness in the spin outcome for any particle must be interpreted as meaning that the various densities continue on along only one path or the other after the measurement. Now, from Eq. (3) we see that every density expression essentially contains a product of the initial and final wavefunctions. It therefore follows that all densities will be zero in a region where either one of these wavefunctions is zero. This fact alerts us to the possibility that we do not need to assume $\psi_i$ collapses in order to have the densities restricted to a single path. The required definiteness will still eventuate as long as the final wavefunction $\psi_f$ is zero along one of the two paths (i.e., if $\psi_f$ overlaps with only one of the branches of $\psi_i$).

On the above basis, our picture for escaping from any undesirable discontinuities is then as follows. As already mentioned, the $+$ and $-$ beams of $\psi_i$ propagate away in two different directions and do not recombine. It is therefore natural to suppose that $\psi_f$, propagating from the future back towards the region of measurement, will be coming from only one of these two different directions. (The alternative would be for the $\psi_f$ of our particle to start as two quite separate wavepackets somewhere in the far future and not combine into one packet until it has travelled back to the time of the present spin measurement. This would be as unlikely in our model as the initial wavefunction $\psi_i$ of a single particle consisting of separate wavepackets throughout the past and not becoming a united wavepacket until the present.) The final wavefunction $\psi_f$ is thus assumed to be zero in one of the two paths (but otherwise arbitrary). To complete the argument, we focus on the branch of $\psi_i$ which does **not** overlap with $\psi_f$ at times after the measurement. This branch will have no further physical effect, since the densities (which constitute the physical reality in this model) are all non-existent along its future. Thus the usual collapse of this portion of $\psi_i$ can be viewed here as the deletion by choice of a branch which is no longer relevant. In particular, choosing to collapse this branch will clearly not entail any discontinuous change in the various densities because they are already zero along this path. The picture formulated here therefore resolves the discontinuity problem outlined earlier, at least for the spin example considered.

---

[17] See, e.g., [20], p. 52.

It should be noted that the magnetic field of the spin apparatus will also cause $\psi_f$ to split into two separate branches as it propagates away from the measurement and on into the past, just as the magnetic field causes $\psi_i$ to split as it propagates forwards in time. This is in accordance with the discussions of Sec. 5 where (as depicted in Figs. 2 and 3) $\psi_i$ tends to spread in the forwards time direction whereas $\psi_f$ tends to spread in the backwards time direction. One of the two branches of $\psi_f$ will overlap with the incoming $\psi_i$ at times earlier than the measurement, in the same way that one of the subsequent two branches of $\psi_i$ overlaps with $\psi_f$ afterwards[18]. (The other branch of $\psi_f$ will propagate away in some other direction and become irrelevant.) This picture is seen to have a natural time symmetry in that the forwards branching of any initial wavefunction is matched by backwards branching of any final wavefunction.

In the above example, the wavefunction of the particle is separated into spatially distinct eigenstates by the measurement process. This, however, is not the only form a measurement may take. For example, the spatial separation may occur in the apparatus wavefunction instead. We will now consider a more general description of measurement, taking the observed system to be an electron, for convenience. The possible measurement outcomes will be taken to be a discrete set of eigenstates $u_n$, with $n = 1, 2, 3, ...$ The electron's initial wavefunction (assuming it is not an eigenstate of the observable quantity to be measured) will be expressible as a superposition of the form $\Sigma c_n u_n$, where each term in this series represents a different measurement outcome and the coefficients $c_n$ satisfy $\Sigma |c_n|^2 = 1$. We will also include in the argument the particular part of the apparatus which indicates the measurement result. For the purposes of illustration, this will be taken to be a pointer reading on a dial. The possible wavefunctions of this pointer which could arise from the measurement will be represented by $p_n$ $(n = 1, 2, 3, ...)$. Now, prior to the measurement interaction taking place, the electron/pointer system may be described by an overall initial wavefunction consisting of the product of the electron's initial wavefunction and the initial wavefunction of the pointer. After the interaction, however, the overall initial wavefunction of the electron/pointer system will have the form of a superposition of product terms:

$$\psi(\mathbf{x}_e, \mathbf{x}_p) = \sum c_n u_n(\mathbf{x}_e) p_n(\mathbf{x}_p) \qquad (25)$$

---

[18] It should be noted that the Schrödinger equation requires at least one branch of $\psi_f$ to overlap with the incoming $\psi_i$ at times before the measurement, given that there is some overlap of $\psi_i$ and $\psi_f$ after the measurement. Similarly, the Schrödinger equation requires that $\psi_f$ must overlap with at least one branch of $\psi_i$ after the measurement, as long as there was non-zero overlap (i.e., the particle existed) before the measurement. This can be seen as follows. As pointed out in Sec. 3, the expressions which the model provides for current density and for the energy-momentum tensor in terms of $\psi_i$ and $\psi_f^*$ satisfy the appropriate conservation equations for charge, energy, etc. This property follows mathematically from $\psi_i$ and $\psi_f$ both satisfying the Schrödinger equation. Since the measurement interaction is being described here in terms of continuous Schrödinger evolution, some non-zero overlap of the system's $\psi_i$ and $\psi_f$ must be maintained through and after the measurement in order for these conserved quantities to continue on. In other words, the Schrödinger equation ensures that the $\psi_i$ and $\psi_f$ in question propagate to some extent in the same regions so that this conservation is achieved.

$\mathbf{x}_e$ and $\mathbf{x}_p$ being the electron and pointer coordinates, respectively. (Here and below, the i for "initial" has been left off each of the wavefunctions to simplify the notation.). Each wavefunction $p_n(\mathbf{x}_p)$ describes a definite and macroscopically distinguishable state of the pointer. This means that these possible wavefunctions for the pointer do not spatially overlap one another. It then follows that the various terms in the summation of Eq. (25) must be separate and non-overlapping in the configuration space in which the overall initial wavefunction $\psi(\mathbf{x}_e, \mathbf{x}_p)$ is defined.

We now proceed in analogy to the argument in the spin one-half example described above. Since the various terms, or "branches", in Eq. (25) are spatially separate and do not recombine, it can be expected that the final wavefunction of the electron/pointer system will be coming from the future along the path of just **one** of these branches of the initial wavefunction (i.e., it will be overlapping with only one branch as it propagates back towards the region of measurement). Hence the various densities defining the physical reality will be zero along all of the branches except one in configuration space. With regard to physical reality, the individual densities for the electron and the pointer in **three-dimensional space** can be extracted from the configuration space description via the procedure outlined in Sec. 7. This is so because the overall wavefunction for the electron/pointer system can be treated similarly to a two-particle wavefunction. The non-overlapped branches in configuration space of the system's initial wavefunction after the measurement are irrelevant and may be ignored. Choosing to delete them obviously does not result in any discontinuous change in the physical densities, because every density is zero already in these regions of configuration space due to the absence of $\psi_f$.

Although the argument above has been framed in terms of the special case of a pointer reading on a dial, it should be clear that it will hold for any type of measurement. This can be seen from the fact that all measurements must involve some form of spatial splitting of the overall particle/apparatus wavefunction into non-overlapping branches in order to make the experimental results identifiable. The above analysis thus provides a way of avoiding the discontinuous changes usually invoked in describing measurements in quantum mechanics.

**Appendix**

To provide justification for the general rule postulated in Sec.8, we will now derive it for a simple case, namely the non-relativistic description of two non-interacting, identical fermions. Referring back to Sec. 7, the "Q density" for two identical particles is given by

$$Q(\mathbf{x},t) = \frac{2}{A}\int \psi_f^*(\mathbf{x},\mathbf{x}',t) Q_\mathbf{x} \psi_i(\mathbf{x},\mathbf{x}',t)\, d^3x' \quad . \tag{26}$$

The further restriction to non-interacting fermions means that the two-particle wavefunctions in this expression must have the antisymmetric forms

$$\psi_i(\mathbf{x},\mathbf{x}',t) = \frac{1}{\sqrt{2}}\left[\psi_{ia}(\mathbf{x},t)\psi_{ib}(\mathbf{x}',t) - \psi_{ia}(\mathbf{x}',t)\psi_{ib}(\mathbf{x},t)\right] \tag{27}$$

$$\psi_f^*(\mathbf{x},\mathbf{x}',t) = \frac{1}{\sqrt{2}}\left[\psi_{fa}^*(\mathbf{x},t)\psi_{fb}^*(\mathbf{x}',t) - \psi_{fa}^*(\mathbf{x}',t)\psi_{fb}^*(\mathbf{x},t)\right] \tag{28}$$

where the single-particle states $\psi_{ia}(\mathbf{x},t)$ and $\psi_{ib}(\mathbf{x},t)$ are initial wavefunctions evolved forward from time $t_1$ and similarly $\psi_{fa}(\mathbf{x},t)$ and $\psi_{fb}(\mathbf{x},t)$ are final wavefunctions evolved backwards from $t_2$. For our purposes, we need to re-express these states in terms of propagators coming from $t_1$ and $t_2$. Adopting the notation

$$(\mathbf{x},t) \equiv x$$

to simplify our expressions, this can be achieved via the following equations:

$$\psi_i(x) = \int K_R(x,x')\psi_i(x')\, d^3x' \qquad (t > t') \tag{29}$$

$$\psi_f^*(x) = -\int K_A^*(x,x')\psi_f^*(x')\, d^3x' \qquad (t < t') \tag{30}$$

where $K_R$ and $K_A$ are the well-known retarded and advanced propagators, respectively. Both $K_R$ and $K_A$ are solutions of[19]

$$\left[H(x) - i\hbar\frac{\partial}{\partial t}\right]K(x,x') = -i\hbar\delta^4(x-x')$$

---

[19] See, e.g., [14], p. 750, or [17], p. 56.

where $H(x)$ is the Schrödinger Hamiltonian operator.

Using the identity

$$K_A{}^*(x,x') = -K_R(x',x)$$

Eq. (30) can be rewritten as

$$\psi_f{}^*(x) = \int \psi_f{}^*(x') K_R(x',x) d^3x' \qquad (t < t') \qquad (31)$$

so that both $\psi_i(x)$ and $\psi_f{}^*(x)$ are now expressed in terms of the retarded propagator. Having eliminated any advanced propagators, we will simplify the notation by dropping the subscript from $K_R$. Eqs. (29) and (31) can now be combined with (27) and (28), respectively, to give

$$\psi_i(x,x') = \frac{1}{\sqrt{2}} \iint \left[ K(x,x_1)K(x',x_3) - K(x',x_1)K(x,x_3) \right] \psi_{ia}(x_1)\psi_{ib}(x_3)\, d^3x_1\, d^3x_3 \qquad (32)$$

$$\psi_f{}^*(x,x') = \frac{1}{\sqrt{2}} \iint \psi_{fa}{}^*(x_2)\psi_{fb}{}^*(x_4) \left[ K(x_2,x)K(x_4,x') - K(x_2,x')K(x_4,x) \right] d^3x_2\, d^3x_4 \qquad (33)$$

with the understanding that, in these two equations, $x$ and $x'$ refer to the same time t:

$$x = (\mathbf{x},t), \qquad x' = (\mathbf{x}',t)$$

and that the various times involved satisfy

$$t_1 < t < t_2$$

$$t_3 < t < t_4$$

Finally, inserting (32) and (33) in (26) and employing the identity

$$\int K(x_2,x') K(x',x_1)\, d^3x' = K(x_2,x_1) \qquad t_1 < t' < t_2 \qquad (34)$$

the "Q density" can be written as

$$Q(x) = \frac{1}{A} \iiiint d^3x_1\, d^3x_2\, d^3x_3\, d^3x_4$$

$$\Big\{ [\psi_{fa}{}^*(x_2) K(x_2,x) Q_x K(x,x_1)\psi_{ia}(x_1)][\psi_{fb}{}^*(x_4) K(x_4,x_3)\psi_{ib}(x_3)]$$

$$+ [\psi_{fa}{}^*(x_2) K(x_2,x_1)\psi_{ia}(x_1)][\psi_{fb}{}^*(x_4) K(x_4,x) Q_x K(x,x_3)\psi_{ib}(x_3)] \qquad (35)$$

$$- [\psi_{fb}{}^*(x_4) K(x_4,x) Q_x K(x,x_1)\psi_{ia}(x_1)][\psi_{fa}{}^*(x_2) K(x_2,x_3)\psi_{ib}(x_3)]$$

$$- [\psi_{fb}{}^*(x_4) K(x_4,x_1)\psi_{ia}(x_1)][\psi_{fa}{}^*(x_2) K(x_2,x)\widehat{Q_x K}(x,x_3)\psi_{ib}(x_3)] \Big\}$$

Having obtained the required density expression, the structure it possesses may be made more transparent by also writing down the corresponding amplitude connecting the initial and final states and then comparing the two expressions. This amplitude has the form

$$A = \iint \psi_f{}^*(x, x') \psi_i(x, x') \, d^3x \, d^3x'$$

which, using (32), (33) and (34), can be written as

$$\begin{aligned}A = \iiiint d^3x_1 \, d^3x_2 \, d^3x_3 \, d^3x_4 \\ \left\{ \left[\psi_{fa}{}^*(x_2) K(x_2, x_1) \psi_{ia}(x_1)\right]\left[\psi_{fb}{}^*(x_4) K(x_4, x_3) \psi_{ib}(x_3)\right] \right. \\ \left. - \left[\psi_{fb}{}^*(x_4) K(x_4, x_1) \psi_{ia}(x_1)\right]\left[\psi_{fa}{}^*(x_2) K(x_2, x_3) \psi_{ib}(x_3)\right] \right\}\end{aligned} \quad (36)$$

Comparing the density expression (35) with this amplitude, the structure of the former is seen to correspond to taking one propagator at a time in the amplitude and making a substitution of the form

$$K(x_2, x_1) \rightarrow \frac{1}{A} K(x_2, x) Q_x K(x, x_1) \quad (37)$$

This correspondence is in agreement with the general prescription proposed in Sec. 8 for evaluating densities (i.e., in agreement with expression (16)). Hence we have succeeded in deriving the desired result.